\documentstyle[12pt,twoside,fleqn,psfig]{article}
\begin{document}\hbadness=10000
\title{$B_c$ Production at RHIC as a Signal for Deconfinement
}
\author{Robert L. Thews and Johann Rafelski\\
\\ 
        Department of Physics,
        University of Arizona\\
        Tucson, AZ 85721, USA}
\date{July 14, 1999}
\maketitle

\begin{abstract}\noindent 
{The $B_c$ meson is the bound state of $b\bar{c}$ (or $\bar{b}c$)
whose recent detection is the first step toward completion of
 the spectroscopy of heavy quark mesonic states.
The  $b$-$c$ states have properties that conveniently fill the gap
between the $J/\psi$ and the  $\Upsilon$ states. Thus it is
probable that at RHIC the $B_c$ mesons will serve
as a probe of deconfined  matter. We find that
significant differences arise for $B_c$ formation in deconfined and
confined matter.  Our initial calculations suggest that:

(a) The rates of normal hadronic production mechanisms at RHIC energies
are {\underbar{not}} sufficient to produce a detectable number of
$B_c$ mesons.

(b) If a region of deconfined quarks and gluons is formed,
the production (and survival) rate can be enhanced by several orders
of magnitude.

(c) The observation of $B_c$ mesons at RHIC would signal a source of
deconfined charmed quarks, and the rate of $B_c$ production will
be a measure of the initial density and temperature of that source.\\

}
\end{abstract}
\section{Introduction}
This work\cite{Thews1} investigates the possibility that
the production of $B_c$ mesons at RHIC may serve as a
signal for the presence (or absence) of a deconfined state of matter
\cite{Thews2}.
The study of the b-c sector has the advantage of a long history of
potential model analysis in the $b\bar{b}$ and $c\bar{c}$ sectors.  These
studies have provided robust predictions for the mass and lifetime of
the $B_c$ states\cite{Thews4}, and the recent measurements 
by CDF\cite{Thews5} are consistent with those calculations.  

First let us estimate at RHIC the production rate of different 
heavy quarks and mesons,  which one would expect if it results just from a
superposition of the initial nucleon-nucleon collisions. 
For heavy quark production, pQCD calculations for p-p interactions
fit present accelerator data and bracket the RHIC energy range.
Hard Probes Collaboration\cite{Thews6}
estimates indicate about 10 $c\bar{c}$
pairs and 0.05 $b\bar{b}$ pairs per central collision at RHIC.   
$J/\psi$ and $\Upsilon$ production involves the use of some model, such as
the Hard Probes color singlet fits\cite{Thews7}, which would predict
bound state fractions of order somewhat less than the one percent level.

A similar analysis for $B_c$ production reaches significantly different 
results\footnote{In the following we include in the term $B_c$ also the vector 
1S state $B_c^*$, since its mass splitting should only allow an
electromagnetic decay into the pseudoscalar ground state and
thus both will contribute identically in the experimental
signatures.} Since the $b$ and $\bar{c}$ must be produced
in the same nucleon-nucleon interaction, parton subprocesses
of order $\alpha_s^4$ are the leading order contributions.  
This leads to a substantial reduction of the bound state fraction
$$R_b\equiv\frac{B_c+B_c^*}{b\bar{b}}$$ 
relative to the few percent levels for the
corresponding $\Upsilon$ state fractions.  At RHIC energies, 
typical values are $R_b=3-10 \times 10^{-5}$, with the uncertainty from 
the scale choice in the pQCD calculations\cite{Thews8}.   

To convert these numbers into $B_c$ production
predictions for RHIC, we have looked at two scenarios for
the luminosity.  a)  The ``first year" case assumes a luminosity of
20 inverse microbarns with no trigger.  b)  The ``design" luminosity
assumes 65 Hz event rate with a 10\% centrality trigger in
Phenix, and uses $10^7$ sec/year.
The predictions we obtain are listed in Table 1.
Included in the
estimates are both the weak branching fraction of the $B_c$ plus
the dimuon decay fraction for $J/\psi$.  Similar numbers are
shown for the $J/\psi$ and $\Upsilon$ production and detection via 
$\mu^+\mu^-$, and also the underlying heavy quark production 
which may be useful to make contact with other estimates.  {\bf One sees
easily that in this scenario there is no hope of seeing $B_c$'s 
at RHIC.} 
\begin{table}[!h]
\caption{RHIC yields for heavy quark systems.}
\begin{center}
\begin{tabular}{|l|cc|} 
\hline\hline
Observable\hfill events&First Year&Design Luminosity\\
\hline
$c\bar c$-pairs & $2.8\,10^8$ & $6.5\,10^9$ \\
$b\bar b$-pairs & $1.2\,10^6$ & $3.2\,10^7$ \\
$J/\Psi\to\mu^+\mu^-$ & $1.6\, 10^5$ & $3.9\,10^6$ \\
$\Upsilon(1s)\to\mu^+\mu^-$ &140 & 3800 \\
\hline\hline
$B_c\stackrel{2.5\%}{\to} J/\psi\it{l}\nu\stackrel{6\%}{\to}\mu^+\mu^-\it{l}\nu$
&&\\
\hline\hline
(No Deconfined Phase) & 0.05--0.18 & 1.5--4.9 \\
\hline
(QGP+$c\bar{c}$ in Chemical Equil.)& 18 & 490 \\
\hline
(Only initial $c\bar{c}$ at $T_o$ = 500 MeV) & 130 & 3530 \\
\hline
(Only initial $c\bar{c}$ at $T_o$ = 400 MeV) & 235 & 6420 \\
\hline
(Only initial $c\bar{c}$ at $T_o$ = 300 MeV) & 475 & 12900 \\
\hline\hline
\end{tabular}
\end{center}
\end{table}

\section{Deconfinement Scenario}

Now the principal reason for our interest - could deconfinement
change the $B_c$ production rate at RHIC?  We have investigated the following
scenario:  In those events in which a $b\bar{b}$ pair are produced, 
one could avoid the small $B_c$ formation fraction if the $b$-quarks
are allowed to form
bound states by combining with $c$-quarks from among
the 10 $c\bar{c}$ pairs already produced by independent
nucleon-nucleon collisions in the same event.  This can occur
if and only if there is a region of deconfinement which allows 
a spatial overlap of the b and c quarks.  
In addition, one would expect some $c\bar{c}$
production in the deconfined phase during its lifetime, as a result of the
approach toward chemical equilibration.   The large binding energy
of $B_c$ (840 Mev) would favor their early ``freezing out"
and they will tend to survive as the temperature drops 
to the phase transition
value.  The same effect for the B mesons and indeed for the $B_s$ will
not be so competitive, since these states are not bound at the 
initial high temperatures (or equivalently they are ionized 
at a relatively high rate by thermal gluons).

To do a quantitative estimate of these effects, 
we calculate the dissociation rate of bound states
due to collisions with gluons, utilizing a quarkonium
break-up cross section based on the operator product expansion
\cite{Thews9}:
\begin{equation}
\sigma_B(k) = {2\pi\over 3} \left ({32\over 3}\right )^2 
\left ({2\mu\over \epsilon_o}\right )^{1/2}
{1\over 4\mu^2} {(k/\epsilon_o - 1)^{3/2}\over (k/\epsilon_o)^5},
\end{equation}
where $k$ is the gluon momentum, $\epsilon_o$ the binding energy,
and $\mu$ the reduced mass of the quarkonium system.  This form
assumes the quarkonium system has a spatial size small compared
with the inverse of $\Lambda_{QCD}$, and its bound state
spectrum is close to that in a nonrelativistic Coulomb potential.
The magnitude of the cross section is controlled just by the
geometric factor ${1\over 4\mu^2}$, and its rate of increase
in the region just above threshold is due to phase space and
the p-wave color dipole interaction.

\begin{center}
\begin{figure}[htb]
\vskip -0.7cm
\hskip 0.5cm\psfig{figure=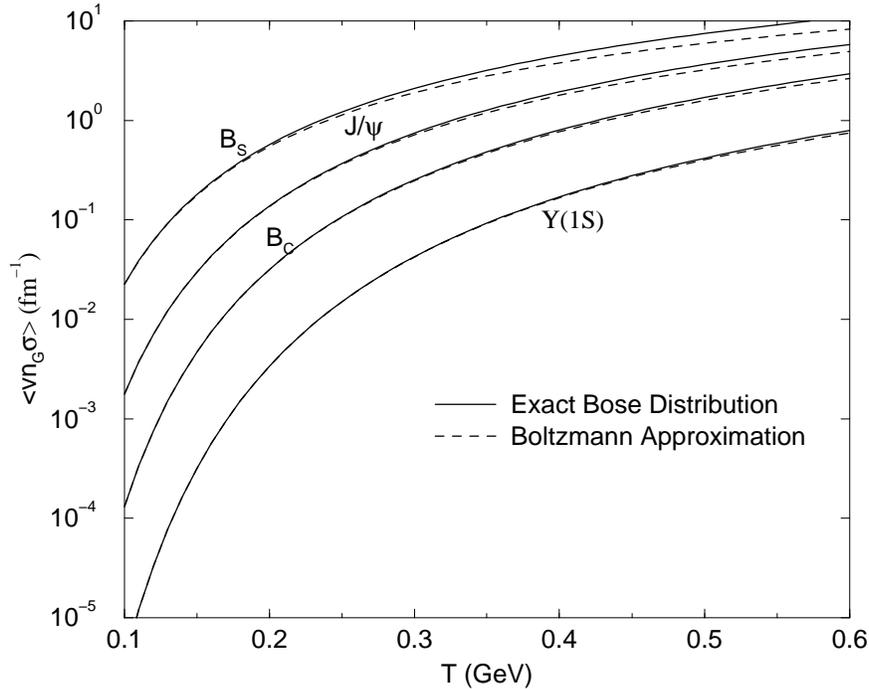,width=4.5in}
\caption{Thermal QGP quarkonium dissociation rates as functions of temperature.
\label{breakuprate}}
\end{figure}
\end{center}

\begin{center}
\begin{figure}[htb]
\vskip -.7cm
\hskip 0.5cm\psfig{figure=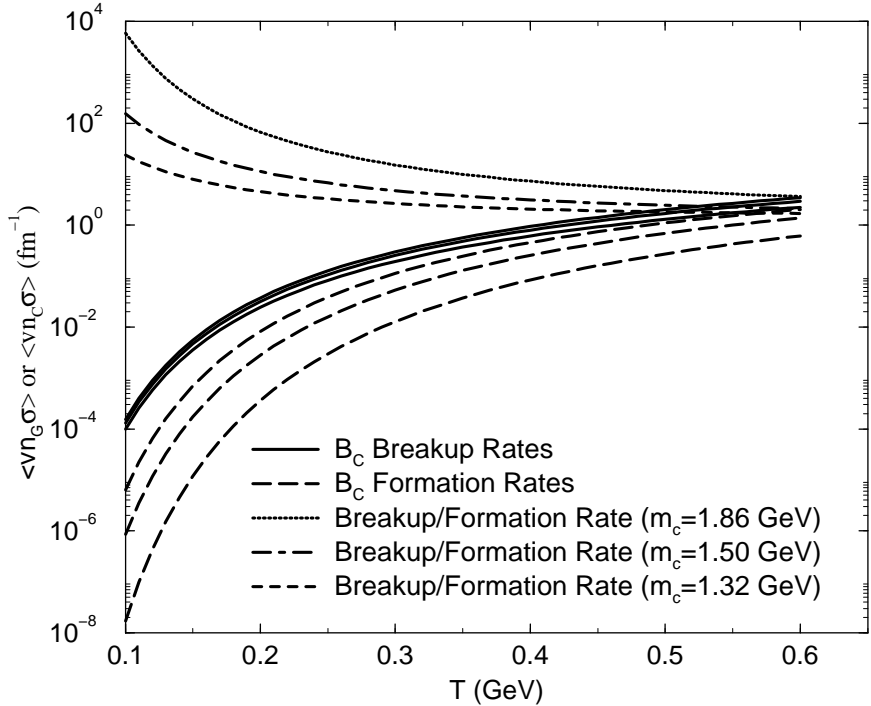,width=4.5in}
\caption{Thermal QGP $B_c$ formation and dissociation rates 
as functions of temperature.\label{thermalrates}}
\end{figure}
\end{center}

For the breakup rate $\lambda_B$ of $B_c$ states in deconfined matter, we
calculate the thermal average:
\begin{equation}
\lambda_B = \left < v_g n_g \sigma_B \right > = {8 \over {\pi^2}}
\int_{\epsilon_o}^{\infty}k^2 dk\ e^{-{k \over T}}\ \sigma_B(k),
\end{equation}
where $v_g$ = 1 and all modes of massless color octet gluons
have been included.  Numerical results for these rates are shown
in Fig. \ref{breakuprate}.  For comparison, breakup rates are also
shown for the $J/\psi$ and $\Upsilon$ (and even the $B_s$, but the
 the approximations made for this  cross section probably have 
a very marginal validity in view of  such
a large state).  One sees that in the range of temperatures expected
at RHIC, these breakup rates 
for $B_c$ lead to time scales of order $1-10\ {\rm fm}$. 
\newpage
For an estimate of the corresponding cross section for the formation
reaction $\sigma_F(b + \bar{c} \to B_c + g)$ we utilized detailed balance 
relations. 
This leads to a finite value of $\sigma_F$ at threshold, since
it is an exothermic reaction.  In the approximation that
the massive $b$-quarks are stationary, which is expected
to be a reasonable approximation due to their energy loss in
the hot plasma \cite{Thews10},  
the formation rate is then calculated 
for a thermal distribution of charm quarks:
\begin{equation}
\lambda_F = \left < v_c n_c \sigma_F \right > = {3 \over {\pi^2}}
\int_{0}^{\infty}\left ({p\over E_p}\right )p^2 dp\ 
 e^{-{ E_p\over T}}\ \sigma_F(p)
\end{equation}         
where $E_p=\sqrt{p^2 + m_c^2}$.
These formation rates are shown in Fig. \ref{thermalrates}.  They
have been calculated for three different values of charm quark
mass. It is apparent that the results are quite sensitive to
this choice, due to the strong dependence of total charm quark
population.  The same values of $m_c$ have very little effect
on the breakup rates, since they only change the overall scale in
the geometric factor of the breakup cross section.

\newpage
Also shown in Fig. \ref{thermalrates} are the ratios 
$\lambda_B/\lambda_F$, which in our normalization is 
related to the bound state fraction in the
equilibrium limit:\footnote{
This bound state fraction is reached if the system has enough time
in its dynamical evolution to relax to the steady-state solution
at each temperature.  We have verified that this is roughly the 
case down to $T = 300$\,MeV, at which point the $B_c$ 
abundance begins to freeze out.}

\begin{equation}\label{Rb}
R_b\equiv {{B_c + B_c^*} \over b\bar{b}} =  {{3 \over 2} {\lambda_F \over \lambda_B} 
 \over {1 + {3 \over 4} {\lambda_F \over \lambda_B}}}\,.
\end{equation}
Note that this ratio approaches its upper limit of 2 when
the formation rate dominates over the breakup rate.  This corresponds
to the situation in which every b-quark produced in the initial
collisions emerges as a $B_c$ bound state. 

We choose a transition temperature $T_f = 160$\,MeV at which to
evaluate the final bound state populations.  Here the equilibrium  
bound state fraction $R_b$ 
drops to as low as several percent, but it is at least a factor of
100 above what one may expect in the no-deconfinement scenario. 
We have chosen to use the equilibrium ratios although at this
final temperature the rates are not sufficient for them to be 
approached.  This provides an even more conservative estimate
for the final bound state populations.
The corresponding entries in the Table for numbers of $B_c$ mesons
(labeled QGP + $c\bar{c}$ in Chemical Equil.)   
uses this conservative lower limit estimate. 

\begin{center}
\begin{figure}[htb]
\hskip 0.5cm\psfig{figure=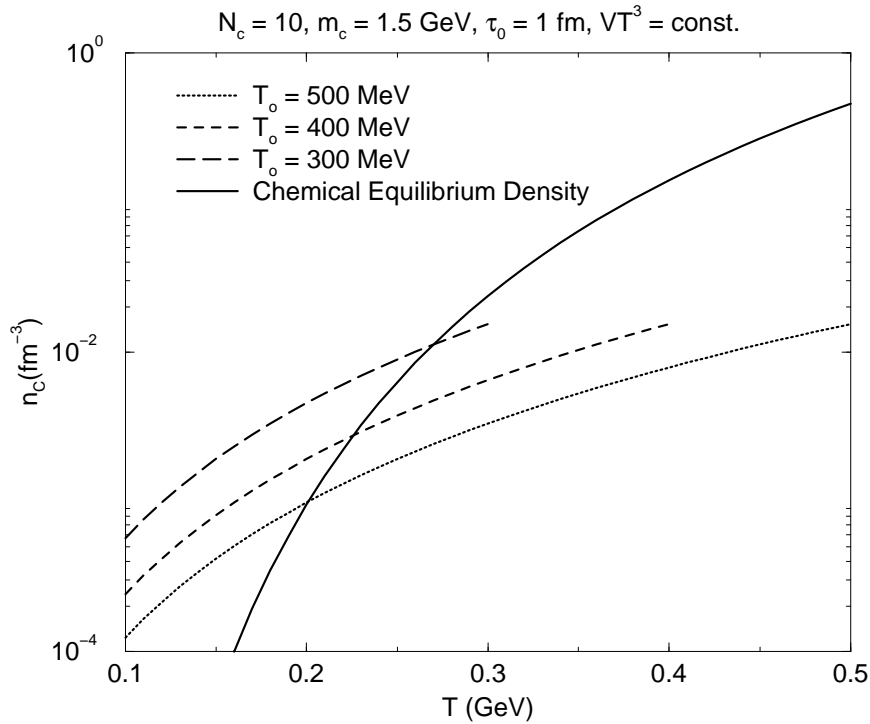,width=4.5in}
\caption{$c$-quark density from initial production at 
RHIC.\label{charmdensity}}
\end{figure}
\end{center}

Implicitly, this analysis
uses the full chemical equilibrium density for c-quarks.  
To get a more
realistic limit we repeated the calculation, using only the
initially-produced $c$-quarks in the formation 
rate.  From the initial population of
10 $c\bar{c}$-pairs produced via nucleon-nucleon collisions in 
a central Au-Au collision at RHIC, and an initial volume 
$V_o = \pi (R_{Au})^2 \tau_o$ with $\tau_o$ = 1.0\ {\rm fm}, one
concludes that only for initial temperatures $T_o < 300$\,MeV 
is the initial charm quark density comparable to that for full
chemical equilibrium.  For initial temperature $T_o$ = 500\,MeV, for
example, the chemical equilibrium charm quark density would be 
about a factor of 40 higher than that 
actually provided by the initially-produced
charm quarks.  As temperature decreases below $T_o$, the isentropic
expansion $VT^3=$\,Const. leads to a decrease in the $c$-quark
density proportional to $T^3$, rather than the 
$e^{-m_c/T}$ of chemical equilibrium.
We have verified that the rates of 
both charm annihilation and production
in a deconfined state for $T < 300$\,MeV then lead to charm
quark occupancies which exceed those for chemical equilibrium as one
approaches the transition point~\cite{Thews11}.
Fig.\,\ref{charmdensity} displays a comparison of chemical
equilibrium charm quark densities and those resulting from a
constant number of initially-produced charm quarks with
isentropic expansion.

\begin{center}
\begin{figure}[htb]
\vskip -0.2cm
\hskip 0.5cm\psfig{figure=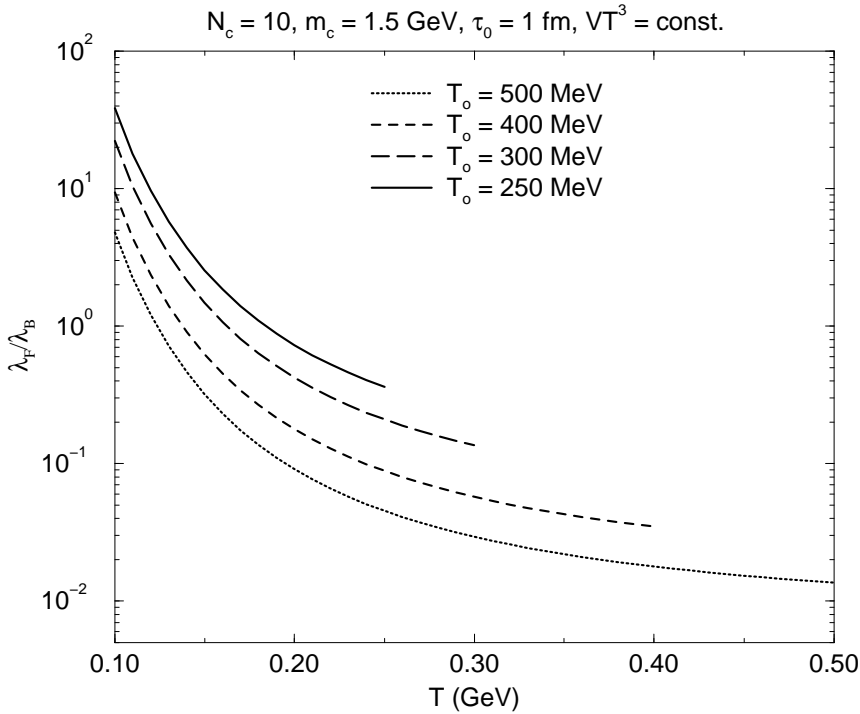,width=4.5in}
\caption{Ratio of formation to break-up rates of $B_c$
as function of temperature for fixed charm quark abundance.
\label{bcfractions}}
\end{figure}
\end{center}

These more realistic charm quark densities are used to recalculate the
formation rates, and the resulting ratios $\lambda_F/\lambda_B$ are 
shown in Fig.\,\ref{bcfractions} for several values of 
initial temperature $T_o$. 
The last few rows in the Table show
the corresponding $B_c$ numbers at RHIC in this scenario, where
we have used the equilibrium bound state fractions again at
a final temperature $T_f$ = 160 MeV.  They
depend quite strongly on the initial temperature, which determines
the final charm density through the assumed isentropic expansion.

We are in the process of refining these preliminary results \cite{Thews11}. 
Initial numerical solutions of the kinetic equations 
using time-dependent formation and breakup rates indicate
the final bound state populations saturate at values appropriate
to those for equilibrium temperatures somewhat above the
transition values.  This would be expected, since the
rates at low temperatures are not sufficient to reach the
equilibrium solutions before the volume expansion 
reduces the temperature to even lower values.  Also, 
production and annihilation of additional charm quark pairs 
is most effective at higher temperatures, which enhances the
effective formation rates.  Both of these effects will enhance
the bound state production fractions for the higher initial
temperatures, and reduce it somewhat for lower initial
temperatures.  However,
it appears that the sensitivity to the parameters of the deconfined
state will remain, making the  $B_c$ signal a sensitive probe
of QGP.

While numerical considerations presented here will see a 
considerable refinement in the near future \cite{Thews11}, the 
firm conclusion we are able to make today is that should QGP
be formed at RHIC there would be a very significant enhancement 
of the formation of $B_c$ mesons which can be observed. 
The primary mechanism responsible for this enhancement is the
interaction of initially-produced bottom and charmed quarks,
which 
will not operate in a confining phase. The observation 
of any $B_c$'s at RHIC is thus both
a ``smoking gun" signal of deconfinement and a probe of the 
initial temperature of the system and the initial density of
deconfined charm.

{\it Acknowledgment}: This work was supported  by a grant from 
the U.S. Department of Energy,  DE-FG03-95ER40937.


\end{document}